\documentclass{acm_proc_article-sp}

\usepackage{graphicx}
\usepackage{epsfig}
\usepackage{amssymb}
\usepackage{amsmath}
\usepackage{hyperref}
\usepackage[T1]{fontenc}
\usepackage{float}
\usepackage{enumitem}
\makeatletter
\def\@copyrightspace{\relax}
\makeatother

\date{}
\begin{document}

\title{LocLinkVis: A Geographic Information Retrieval-Based System for Large-Scale Exploratory Search}

%
%
%
%
%

\numberofauthors{3} 
%
\author{
%
%
\alignauthor Alex Olieman\\
       \affaddr{University of Amsterdam}\\
       \email{olieman@uva.nl}
\alignauthor Jaap Kamps\\
       \affaddr{University of Amsterdam}\\
       \email{kamps@uva.nl}
\alignauthor Rosa Merino Claros\\
       \affaddr{University of Amsterdam}\\
       \email{r.merinoclaros@uva.nl}
}

\maketitle

\begin{abstract}
In this paper we present LocLinkVis (Locate-Link-Visualize); a system which supports exploratory information access to a document collection based on geo-referencing and visualization. It uses a gazetteer which contains representations of places ranging from countries to buildings, and that is used to recognize toponyms, disambiguate them into places, and to visualize the resulting spatial footprints.
\end{abstract}


\section{Introduction}

The present day Web search interaction model of typing a simple keyword query and glancing over a list of 10 blue links has not satisfied the needs of (re)searchers who deal with large (thematic) collections. In the cultural heritage and legal domains, for instance, information needs that cannot be answered by any single document, but require comparing several documents or an overview of documents that satisfy the query, are quite common.

One important direction for improving access to sufficiently large collections, such as those curated by archives and libraries, is geo-referencing: linking information to geographical location \cite{Clough2011}. Links from documents to geospatial entities can be indexed by a search system, allowing users to freely explore the geographical aspect of the collection, or to view the results of a thematic query on a map. This is an important next step in making digital cultural heritage collections more accessible to scholars in various branches of the humanities \cite{Simon2015}. Furthermore, the most common types of interaction that today's users have with digital maps, \emph{pan} and \emph{zoom}, translate exceptionally well to existing exploratory search strategies: \emph{shift} and \emph{narrow} \cite{Samp2014}.

Existing Geographic Information Retrieval (GIR) systems designed for archives, however, are limited by their representation of locations. These systems represent \emph{place} as \emph{point} (i.e. a single latitude-longitude pair) \cite{Borin2014,Clough2011}, which uncontestedly simplifies the technical requirements of the system, but also limits the possibilities for the meaningful visualization of document results. We argue that there is more to visualizing a mentioned location than a single point. Qualitatively, it can make a big difference whether someone speaks about a country, region, municipality, metropolis, highway, town, square, street, school, restaurant, or monument. How can this difference be shown in the visualization? We propose a deceptively simple answer to this question: by displaying the same geometry that is used to represent these locations on a map.

In this paper we present LocLinkVis; a system which supports exploratory information access to a document collection based on geo-referencing and visualization. It uses a gazetteer which contains representations of places ranging from countries to buildings, and which is used to recognize toponyms, disambiguate them into places, and to visualize the resulting spatial footprints. OpenStreetMap\footnote{\url{http://www.openstreetmap.org/}} (OSM) is the only data source that is needed to fill the gazetteer.

Our current LocLinkVis prototype covers the Netherlands and neighboring regions, but it can easily be populated by arbitrary subsets of OSM data (which has planet-wide coverage\footnote{See \url{https://tyrasd.github.io/osm-node-density/}for OSM's coverage in 2014}). The current prototype runs comfortably on a commodity laptop,  but would benefit from more powerful hardware if it were to provide global coverage. For the document-side of the system our proof-of-concept focuses on existing digital collections of parliamentary proceedings from Canada, the Netherlands, and the United Kingdom. The current LocLinkVis prototype is limited to a Dutch gazetteer and corpus, but will soon be expanded with the Canadian and UK corpora to match the coverage of the general PoliticalMashup\footnote{\url{http://search.politicalmashup.nl}} search engine.

\section{System Architecture}
\begin{figure*}[t]
    \centering
    \includegraphics[width=\textwidth]{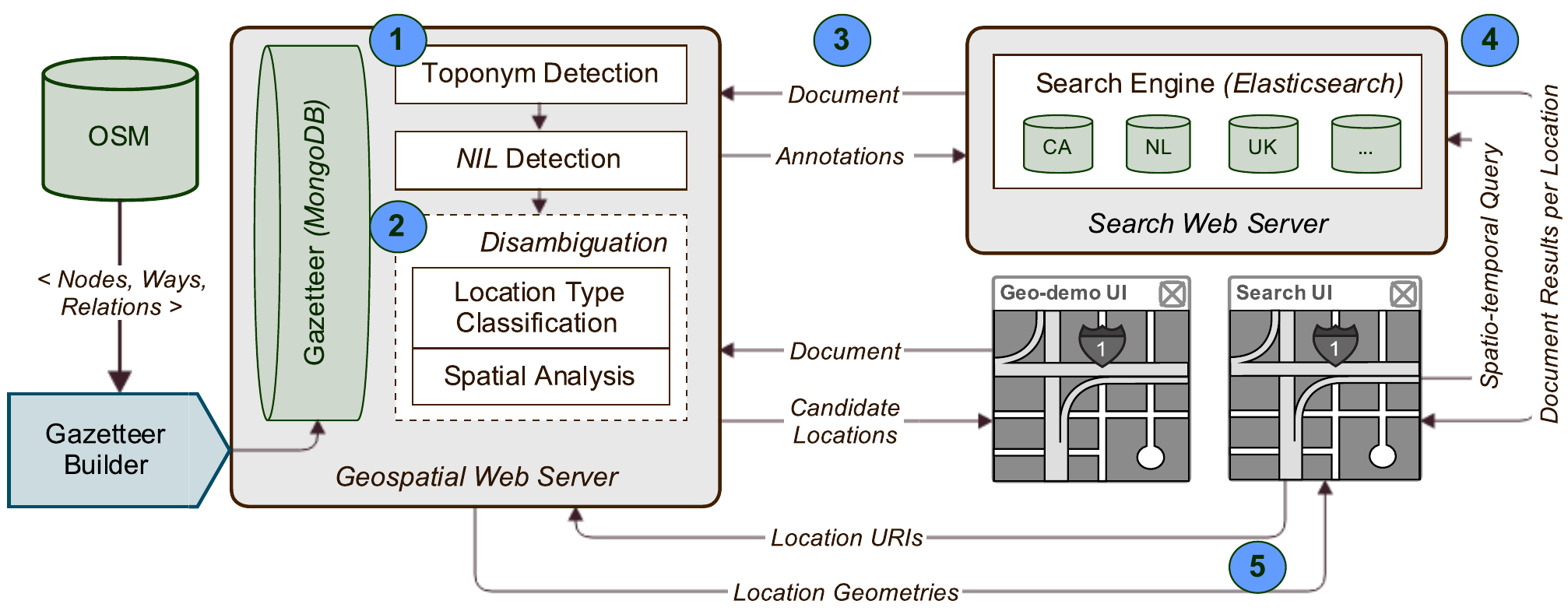}
    \caption{Schematic overview of LocLinkVis' system architecture.}
    \label{fig:figure1}
\end{figure*}
The systems design of LocLinkVis originates from the following requirements. The system should efficiently:
\begin{enumerate}[noitemsep,nolistsep]
\item \emph{detect toponyms} ranging from countries to buildings;
\item \emph{disambiguate} which geographical features the toponyms refer to;
\item build an \emph{index} of documents by the locations they mention;
\item use a map interface's \emph{viewport} and \emph{zoom level} as query parameters;
\item \emph{visualize} results on a map, at a suitable level of abstraction.
\end{enumerate}
Figure 1 shows the components of the resulting system, and which communication occurs between them. The numbers in this schematic indicate which requirement is satisfied in (or between) which part(s) of the system.

The \textbf{gazetteer builder} takes OSM data as input, in the form of efficient binary dumps as well as incremental XML updates. Its primary purpose is to transform OSM's normalized (i.e. with internal references) \texttt{Node}, \texttt{Way}, and \texttt{Relation} instances to denormalized GeoJSON\footnote{\url{http://geojson.org/geojson-spec.html}} features. This JSON-based format for geospatial data structures is better suited towards retrieving and rendering spatial data, whereas OSM's data schema favors editing it. Second, while the OSM data is being imported, instances of the same geographical type (e.g. city or road) that share a toponym and geometrically intersect, are merged into a single gazetteer feature. OSM contains data at fine level of granularity, e.g. to indicate that a particular road section goes through a tunnel. Our aim is to simplify (the side-effects of) such details when they are undesirable for spatial document search.

Our \textbf{gazetteer}, like its paper media predecessors, provides an easily searchable list of toponyms, each accompanied by facts about its location amongst other aspects. Modern gazetteers, such as the Getty Thesaurus of Geographic Names\footnote{\url{http://www.getty.edu/research/tools/vocabularies/tgn/}}, commonly represent the location of places as coordinate pairs. With thanks to the OSM contributors, we can now provide detailed geometries for the features in our gazetteer. Besides the data that is copied and denormalized from OSM, we add to each feature the identifiers of the OSM instances it was derived from, and build an index of features by all their known names (including historic toponyms).

\begin{figure*}[t]
    \centering
    \includegraphics[width=\textwidth]{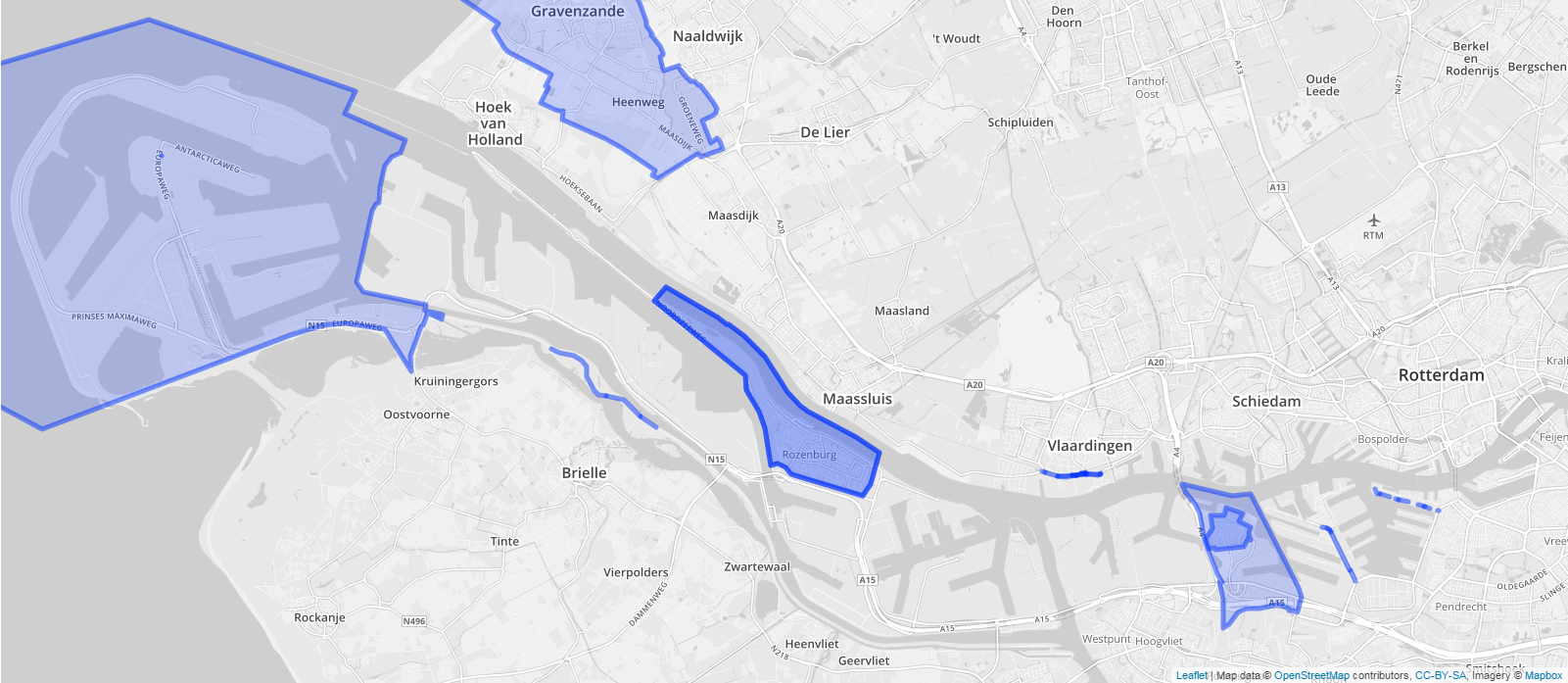}
    \caption{Slippy map with highlighted places that are mentioned in context of the Rotterdam harbor.}
    \label{fig:figure2}
\end{figure*}
\begin{figure*}[t]
    \centering
    \includegraphics[width=\textwidth]{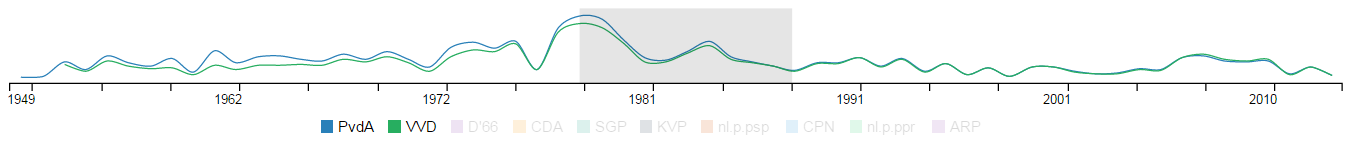}
    \caption{Timeline widget showing the number of results for current Dutch government parties, with a selection of the period 1978-1987.}
    \label{fig:figure3}
\end{figure*}

\textbf{Toponym detection} entails identifying which tokens from an input document refer to a place. LocLinkVis takes a lexicon-based approach in which all known toponyms from the gazetteer are encoded into a finite state automaton by the Aho-Corasick algorithm \cite{Aho1975}. This allows for efficient string matching between an input document and a large set of toponyms. At this stage the detected phrases are candidate toponyms, which may or may not refer to a gazetteer feature.

\textbf{\emph{NIL} detection} is the process by which candidate toponyms may be identified as not-referring to any location that is represented in the gazetteer. We opted for a corpus-independent approach, which starts by counting the frequency with which toponyms from the gazetteer are mentioned in a Wiktionary\footnote{\url{https://www.wiktionary.org/}} dump of the desired language. Wiktionary, for this purpose, serves as a compact, but high-coverage, cross section of lexical forms within a language. Possible toponyms that are mentioned frequently, it seems safe to assume, refer either to well-known places or are homographs of common words or other names. LocLinkVis uses Support Vector Machines that have been trained on a small number of training examples to make this distinction. Besides Wiktionary occurrence, the feature space that is used for classification includes variables obtained from SPARQL queries to DBpedia, as well as queries to the gazetteer.

\textbf{Location type classification} takes the lexical context of mentioned toponyms into account for disambiguation. It computes for any toponym (not limited to our gazetteer) the probabilities that it refers to an instance of the types \texttt{Country}, \texttt{Province}, (body of) \texttt{Water}, \texttt{Municipality}, \texttt{Neighborhood}, \texttt{Road}, and \texttt{Building}. It does so on the basis of a window of five tokens preceding and following the toponym in question. Numerical tokens are replaced by a normalized form. The same is done for other toponyms within the token window in the training examples, and for unambiguous toponyms in unseen input. LocLinkVis offers the choice between a Naive Bayes and a Maximum Entropy classifier, and uses between 25 and 85 training examples per class.

\textbf{Spatial analysis} is useful to operationalize Tobler's \emph{``first law of geography,''} that \emph{``everything is related to everything else, but near things are more related than distant things.''} \cite{Miller2004} By taking into account the spatial distance between candidate entities, we investigate the hypothesis that nearer features are more frequently mentioned together than distant features. A relatedness measure such as distance can be used with existing methods to collectively rank candidate entities \cite{Shen2014}. We employ a graph-based method, using random walkers to traverse a fully-connected weighted graph of features, to rank candidates. The next step, at the time of writing, is to explore how to integrate the spatial and lexical parts of our disambiguation approach.

The \textbf{geo-demo user interface} (UI), available at\\ \url{http://geodemo.politicalmashup.nl}, allows users to interactively validate our geo-referencing approach. Using this UI, arbitrary text can be submitted to the geospatial web server to view the intermediate results at each stage of the geo-referencing process. It shows which toponyms have been detected, which are thought not to be represented in the gazetteer, and the (detailed) output of location type classification and spatial analysis for all candidate features. A \emph{slippy map}\footnote{\url{http://wiki.openstreetmap.org/wiki/Slippy_Map}} covers the screen, with input fields and geo-referencing details being available from an expandable menu.

The \textbf{search engine} provides indexed access to one or more document collections. In our particular case these are corpora of national parliamentary proceedings in the rich PoliticalMashup XML format, but plain text documents would also be sufficient. In order to enrich a document with geospatial annotations, the text it contains is submitted to the geospatial web server, which returns annotations of the form 
$$\texttt{<\text{Feature URI}, \text{Span}, \text{Confidence}, \text{Bounding Box}>.}$$
A geospatial index ensures that the search engine can be queried with an arbitrary bounding box, and return documents from which any of the annotated locations intersect with the bounding box. If the documents also feature a temporal property (e.g. creation date or period of activity), it can enable further meaningful filtering of the results.

LocLinkVis' \textbf{search UI} is described in the following section.

\section{Search UI Design}
The main aim of our search UI is to provide an interactive geospatial overview of document results. In contrast to traditional GIR systems, which have focused on associating documents with a single location (often in the form of a point), this UI displays all places that are mentioned in the documents (see Figure 2), and with the same fidelity to which they are rendered on the background map. To assist in searching document collections with an interesting temporal aspect, we have added a timeline widget which further filters the result set.

The timeline widget, see Figure 3, shows the number of places that are mentioned in the result set and intersect with the map viewport. A categorical variable (i.e. \emph{facet}) can be used to break the total frequency down into result categories. The user may make a selection on this timeline, which acts as a temporal filter within the larger query. Additionally, besides manual adjustment of the selected time period, controls are provided which enable moving window playback which shows the dynamic of how often places were mentioned through time.

By these means the search UI provides a bidirectional binding between the spatial and temporal aspects for the exploration of a search result set. When a user interacts with the slippy map (i.e. by panning and zooming), the chart in the timeline widget is updated to reflect the new viewport. \emph{Vice versa}, when a time period is selected or adjusted, the search result set is (re-)filtered and the highlighted locations on the map are updated accordingly. This enables the user to seamlessly move between shift and narrow search strategies, as well as between the spatial and temporal dimensions of the result space.

The search UI is under active (iterative) development at the time of writing. Hence, our prototype at \url{http://geosearch.politicalmashup.nl} will be changed periodically and may not always be available. We are, for instance, currently experimenting with the placement of result snippets in the interface. When a user clicks a highlighted location, document results from a query could either be displayed on or near the clicked location, or be shown within a fixed screen area. 

\section{Conclusion and Outlook}

In this paper a novel GIR-based system was presented that supports exploratory information access to a document collection by means of geo-referencing and interactive visualization. LocLinkVis distinguishes itself from existing systems that have similar goals by working with a broader range of location types, and by representing a location's geometry with the same fidelity as to which it is rendered on OpenStreetMap. It also features a UI which allows the user to seamlessly move between the shift and narrow search strategies, as well as between the spatial and temporal dimensions of the result space.

This work is situated in the context of a project which aims to improve (exploratory) information access to parliamentary proceedings and related cultural heritage collections. We are working towards a \emph{living lab} with high-quality, reliable prototype systems and ubiquitous interaction logging. This approach allows us to use a combination of (remote) laboratory experiments and analysis of in-the-wild usage to investigate how we can best assist historians, political scientists, journalists, and media scholars in their complex search and sensemaking activities. 

Our current effort to provide accurate results for historical collections mainly consists of collecting historical place names from OSM, e.g. to compensate for spelling variants in older texts. The gazetteer, however, does not yet contain information about during which time period a place was known by a particular name, and only contains the most recent geometric representation that is available. This can lead to a distorted picture e.g. when borders have changed much, or when a monument has not always been in its present-day location. At least for Dutch municipalities the information needed to display the historical situation accurately is available \cite{Zandhuis2014}, and we plan to adapt the gazetteer builder to incorporate it.

\textbf{Acknowledgements} This research was supported by the Netherlands Organization for Scientific Research (ExPoSe project, NWO CI \# 314.99.108; DiLiPaD project, NWO Digging into Data \# 600.006.014). We would like to thank the anonymous reviewers of this paper for their useful remarks.

%
\bibliographystyle{abbrv}
\bibliography{sigproc}  

\begin{thebibliography}{1}

\bibitem{Aho1975}
A.~V. Aho and M.~J. Corasick.
\newblock {Efficient string matching: an aid to bibliographic search}.
\newblock {\em Communications of the ACM}, 18:333--340, 1975.

\bibitem{Borin2014}
L.~Borin, D.~Dannells, and L.-J. Olsson.
\newblock {Geographic visualization of place names in Swedish literary texts}.
\newblock {\em Literary and Linguistic Computing}, 29(3):400--404, May 2014.

\bibitem{Clough2011}
P.~Clough, J.~Tang, M.~M. Hall, and A.~Warner.
\newblock {Linking archival data to location: a case study at the UK National
  Archives}.
\newblock {\em Aslib Proceedings}, 63(2/3):127--147, 2011.

\bibitem{Miller2004}
H.~J. Miller.
\newblock {Tobler's First Law and Spatial Analysis}.
\newblock {\em Annals of the Association of American Geographers},
  94(2):284--289, 2004.

\bibitem{Samp2014}
K.~Samp, C.~Bezuit, and J.~Schneider.
\newblock {Unifying the Shift and Narrow Strategies in Focus + Context
  Exploratory Search}.
\newblock In {\em Proc. of SIGDOC 2014}, Colorado Springs, CO, USA, 2014. ACM.

\bibitem{Shen2014}
W.~Shen, J.~Wang, and J.~Han.
\newblock {Entity Linking with a Knowledge Base: Issues, Techniques, and
  Solutions}.
\newblock {\em IEEE Transactions on Knowledge and Data Engineering},
  4347(2):443--460, 2014.

\bibitem{Simon2015}
R.~Simon, E.~Barker, L.~Isaksen, and P.~de~Soto~Ca{\~n}amares.
\newblock Linking early geospatial documents, one place at a time: Annotation
  of geographic documents with recogito.
\newblock {\em e-Perimetron}, 10(2):49--59, 2015.

\bibitem{Zandhuis2014}
I.~Zandhuis, M.~den Engelse, and E.~{Mac Gillavry}.
\newblock {Dutch historical toponyms in the Semantic Web}.
\newblock In {\em Proc. Workshop Population Reconstruction}, 2014.

\end{thebibliography}
%
%

\end{document}